\documentclass[preprintnumbers,twocolumn,prl,aps,superscriptaddress,footinbib,amsfonts,amsmath,amssymb,bm,showpacs,floatfix]{revtex4-1}

\makeatletter
\renewcommand*{\@fnsymbol}[1]{\ensuremath{\ifcase#1\or *\or \dagger\or \ddagger\or
   \mathsection\or \mathparagraph\or \|\or **\or \dagger\dagger 
   \or \ddagger\ddagger 
   \or \mathsection\mathsection
   \or \mathparagraph\mathparagraph
   \or \|\|
   \else\@ctrerr\fi}}
\makeatother

\usepackage[colorlinks=true,linkcolor=blue,citecolor=blue,urlcolor=blue]{hyperref}

\usepackage[T1]{fontenc}
\usepackage{blindtext}
\usepackage{amsmath}
\usepackage{amssymb}
\usepackage{graphics,graphicx}
\usepackage{dcolumn}
\usepackage{bm}
\usepackage{epsfig}
\usepackage[normalem]{ulem}
\usepackage[usenames]{color}
\usepackage{mathbbol}
\usepackage{epstopdf}
\usepackage{simplewick}
\usepackage[utf8]{inputenc} 
\usepackage{slashed}
\usepackage{soul}
\usepackage[dvipsnames]{xcolor}
\usepackage[export]{adjustbox}
\usepackage{mathtools}

\renewcommand{\texttt}[1]{#1}

\allowdisplaybreaks

\newcommand{\fref}[1]{Fig.~\ref{f.#1}}

\newcommand{\eref}[1]{Eq.~(\ref{e.#1})}

\newcommand{\df}{\mathrm{d}}

\newcommand\muQ{\mu}

\begin{document}

\preprint{JLAB-THY-25-4573}
\preprint{IPARCOS-UCM-25-051}  

\title{First simultaneous analysis of transverse momentum dependent \\ and collinear parton distributions in the proton}

\newcommand*{\ANL}{Physics Division, Argonne National Laboratory, Lemont, Illinois 60439, USA}\affiliation{\ANL}
\newcommand*{\PSU}{Division of Science, Penn State University Berks, Reading, Pennsylvania 19610, USA}\affiliation{\PSU}
\newcommand*{\JLAB}{Jefferson Lab, Newport News, Virginia 23606, USA}\affiliation{\JLAB}
\newcommand*{\WM}{Department of Physics, William \& Mary, Williamsburg, Virginia 23187, USA}\affiliation{\WM}
\newcommand*{\LVC}{Department of Physics, Lebanon Valley College, Annville, Pennsylvania 17003, USA}\affiliation{\LVC}
\newcommand*{\UCM}{Departamento de Física Teórica and IPARCOS, Universidad Complutense de Madrid, \\ Plaza de Ciencias 1, 28040 Madrid, Spain}\affiliation{\UCM}
\newcommand*{\ODU}{Department of Physics, Old Dominion University, Norfolk, Virginia 23529, USA\\
        \vspace*{0.2cm}
        {\bf JAM Collaboration \\ {\footnotesize \ (TMD Analysis Group)}
        \vspace*{0.2cm} }}\affiliation{\ODU}

\author{P.~C.~Barry}
\thanks{\href{mailto:barryp@jlab.org}{barryp@jlab.org}}
\altaffiliation{Present address: Jefferson Lab,
Newport News, Virginia 23606, USA}
\affiliation{\ANL}

\author{A.~Prokudin}
\thanks{\href{mailto:prokudin@jlab.org}{prokudin@jlab.org\\}}
\affiliation{\PSU}\affiliation{\JLAB}

\author{T.~Anderson}
\affiliation{\JLAB}\affiliation{\WM}

\author{C.~Cocuzza}
\affiliation{\WM}

\author{L.~Gamberg}
\affiliation{\PSU}

\author{W.~Melnitchouk}
\affiliation{\JLAB}

\author{E.~Moffat}
\affiliation{\ANL}

\author{D.~Pitonyak}
\affiliation{\LVC}

\author{J.-W.~Qiu}
\affiliation{\JLAB}\affiliation{\WM}

\author{N.~Sato}
\affiliation{\JLAB}

\author{A.~Vladimirov}
\affiliation{\UCM}

\author{R.~M.~Whitehill}
\affiliation{\ODU}

\begin{abstract}
We present the first simultaneous global QCD analysis of unpolarized transverse momentum dependent (TMD) and collinear parton distribution functions (PDFs) in the proton. Our study incorporates data from deep-inelastic scattering, Drell-Yan, inclusive weak boson, $W$+\,charm, and jet production involving PDFs, as well as TMD Drell-Yan and $Z$-boson production data from fixed target and collider experiments sensitive to both TMD and collinear distributions. The analysis is performed at next-to-next-to-leading logarithmic (NNLL) accuracy for QCD resummation in TMD observables and next-to-leading order (NLO) for observables described in collinear factorization. The combined analysis improves knowledge of both TMD and collinear PDFs, particularly in the sea-quark sector, providing a consistent simultaneous description of the aforementioned observables.
\end{abstract}

\maketitle 

{\it Introduction --- }
Over the past half century significant effort has been devoted to studying the internal structure of nucleons in terms of their fundamental quark and gluon constituents within the theory of Quantum Chromodynamics~(QCD)~\cite{Gross:2022hyw}.
Probing the nucleon’s internal structure relies on experimental and phenomenological input, employing QCD factorization theorems \cite{Altarelli:1977zs, Dokshitzer:1977sg, Collins:1984kg, Collins:1987pm, Catani:1998nv, Collins:2011zzd} to separate observables into perturbative hard parts and nonperturbative functions encoding structure information.
In addition, evolution equations derived from factorization theorems 
resum large process-independent logarithmic contributions from the collision-induced radiation and connect observables at different scales.
Accordingly, measurements from fixed target and collider experiments are synergistic, together enabling detailed studies of perturbative and nonperturbative QCD dynamics.

The most well-known nonperturbative functions are the unpolarized collinear parton distribution functions (PDFs)~\cite{Alekhin:2017kpj, Hou:2019efy, Bailey:2020ooq, Ethier:2020way, NNPDF:2021njg, PDF4LHCWorkingGroup:2022cjn, Anderson:2024evk, Cocuzza:2026zoy}. 
In recent years there has been increased focus on the nucleon's three-dimensional structure, encoded in transverse momentum dependent (TMD) parton distribution functions \cite{Kotzinian:1994dv, Tangerman:1994eh, Mulders:1995dh, Boer:1997nt,Boussarie:2023izj} (referred to as ``TMDs'' in this work).  
While PDFs depend only on the longitudinal parton momentum fraction $x$, TMDs are in addition sensitive to the parton's transverse momentum $k_T \equiv |\bm{k_T}|$. 
The study of TMDs has yielded many phenomenological successes, where key nonperturbative inputs to the unpolarized and polarized TMDs and the Collins–Soper (CS) kernel have been inferred~\cite{Bacchetta:2017gcc, Bacchetta:2019sam, Bertone:2019nxa, Scimemi:2019cmh, Bacchetta:2022awv, Wang:2017zym, Vladimirov:2019bfa, Cerutti:2022lmb, Bury:2022czx, Aslan:2024nqg, Bacchetta:2024qre, Bacchetta:2024yzl, Yang:2024drd, Barry:2023qqh, Moos:2020wvd, Moos:2023yfa, Moos:2025sal, Bacchetta:2025ara, Rossi:2025pwh, Vladimirov:2025qrh}.

The main motivation for this work is based on the connection between PDFs and TMDs.
The latter can be expressed in the space of $b_T\equiv |\bm{b}_T|$, the Fourier conjugate to $\bm{k}_T$, $\tilde{f}(x,b_T)= \int \df^2\bm{k}_T\, e^{-i \bm{b}_T \cdot \bm{k}_T} f(x,\bm{k}_T)$.
While TMD and collinear factorizations are formally distinct, the TMD operator for $\tilde{f}(x,b_T)$ matches onto the operator for the PDF $f(x)$ in the small-$b_T$ limit, as formalized by the operator product expansion (OPE) \cite{Collins:1984kg, Collins:2011zzd, Collins:2014jpa, Echevarria:2016scs, Luo:2020epw, Ebert:2020yqt, Ebert:2020qef, Moos:2020wvd} and by integral relations~\cite{Ebert:2022cku, Gonzalez-Hernandez:2023iso, delRio:2024vvq}, which enables explicit construction of the TMDs from collinear PDFs.

In this Letter, we perform the first simultaneous global QCD analysis of collinear PDFs and TMDs of the proton utilizing data from deep-inelastic scattering (DIS), Drell-Yan (DY), inclusive weak boson, $W$+\,charm, and jet production, as well as TMD-sensitive DY and $Z$-boson production data from fixed target and collider experiments.
This work establishes a new paradigm in which collinear PDFs and TMDs are extracted simultaneously, allowing for improved determination of these universal QCD functions and a more consistent theoretical description of observables sensitive to PDFs and TMDs.

{\it Theoretical formalism --- }%
TMD factorization~\cite{Collins:1981uk, Collins:1981va, Collins:1984kg, Ji:2004xq, Becher:2010tm, Collins:2011zzd, Echevarria:2011epo} describes the DY process, $h_1 h_2 \to \ell^+ \ell^- X$, in the region of small transverse momentum $q_T$ of the produced lepton pair relative to its invariant mass~$Q$.
In this regime, the measured cross section has been described in the Collins-Soper-Sterman (CSS) formalism~\cite{Collins:1984kg} and recently presented in terms of TMDs~\cite{Mulders:1995dh, Arnold:2008kf, Collins:2011zzd, Boussarie:2023izj}.
The cross section for proton--proton collisions differential in $Q^2$, rapidity $y$ of the lepton pair, and $q_T$ is given~by  
\begin{align}
 \frac{\df^3\sigma}{\df Q^2\, \df y\, \df q_T^2} 
 & = \sigma_0 \mathcal{P} \sum_{q} c_q^2(Q)\, \mathcal{H}_{q\bar q}^{{\mbox{\tiny \rm DY}}}(Q,\muQ) \int\! \frac{\df^2{\bm b}_T}{(2\pi)^2} \, e^{i {\bm b}_T \cdot {\bm q}_T}
   \nonumber \\
  &
   \hspace*{0.3cm}\times  \tilde f_{q/p}(x_1, b_T; \muQ, \zeta) \,
   \tilde f_{\bar q/p}(x_2, b_T; \muQ, \zeta)
\label{e.DYpTxsec}
\,,
\end{align}
where the sum over $q$ runs over all quarks and antiquarks with electroweak charges $c_q^2$.
Here, $\mathcal{P}$ is the fiducial factor that takes into account experimental cuts on the momenta of the detected leptons~\cite{Scimemi:2019cmh, Bacchetta:2019sam}, 
\mbox{$\sigma_0 \equiv 4\pi^2 \alpha^2/9 Q^2 s$} with $s$ the center of mass energy squared of the reaction, $\mathcal{H}^{{\mbox{\tiny \rm DY}}}_{q\bar q}$ is the hard coefficient function, and \mbox{$x_{1(2)} = (Q/\sqrt{s})\, e^{+(-)y}$} are the partonic momentum fractions.
Two scales, $\mu$ and $\zeta$, regularize the ultraviolet and rapidity divergences, respectively, and are conventionally chosen to be $\mu^2=\zeta=Q^2$ to optimize the perturbative convergence.
Each TMD satisfies the CS equations~\cite{Collins:2017oxh,Boussarie:2023izj}, which govern the dependence on these scales, and several frameworks exist for implementing their solutions --- see Chapter 4 of Ref.~\cite{Boussarie:2023izj} and references therein.

In our study we utilize the $\zeta$-prescription~\cite{Scimemi:2018xaf, Vladimirov:2019bfa} for TMD evolution.
A defining feature of the $\zeta$-prescription is that it uses the null-evolution line ($\mu$, $\zeta_\mu(b_T)$) as the reference scale, defined by
\begin{equation}
\Gamma(\mu) \ln\bigg(\frac{\mu^2}{\zeta_\mu(b_T)}\bigg) - \gamma_V(\mu) 
= 2 \mathcal{D}(b_T;\mu) \frac{\df \ln \zeta_\mu(b_T)}{\df \ln \mu^2},
\label{e.zeta}
\end{equation}
where $\gamma_V$ is the anomalous dimension of the vector form factor~\cite{Moch:2005id, Collins:2017oxh}, and $\Gamma$ is the cusp anomalous dimension~\cite{Polyakov:1980ca, Korchemsky:1985xj, Korchemsky:1987wg, Collins:1989bt, Moch:2004pa, Henn:2019swt}. 
The CS kernel $\mathcal{D}$~\cite{Collins:2011zzd, Collins:2014jpa, Moos:2023yfa, Boussarie:2023izj} determines the rapidity scale evolution of the TMD,
\begin{equation}
\mathcal{D}(b_T;\mu) \!=\! \mathcal{D}_{\rm pert}(b^*;\mu^*)  + 
 \int_{\mu^*}^\mu \frac{\df \mu^\prime}{\mu^\prime}\Gamma(\mu^\prime) + \mathcal{D}_{\rm NP}(b_T).
\label{e.cs}
\end{equation}
Here, $\mathcal{D}_{\rm pert}$ is a perturbatively calculable component that describes the small-$b_T$ behavior~\cite{Vladimirov:2016dll, Moult:2022xzt, Duhr:2022yyp}, and $\mathcal{D}_{\rm NP}$ is a nonperturbative component which dominates the large-$b_T$ region.
The matching between the small- and large-$b_T$ regions is controlled via $b^*$, where $b^*(b_T) = b_T/\sqrt{1+b_T^2/B_{\rm NP}^2}$ with fixed $B_{\rm NP} = 1.5$~GeV$^{-1}$ and $\mu^* = 2 e^{-\gamma_E}/b^*(b_T)$. 
The nonperturbative component is extracted from the data, as discussed below.

In the $\zeta$-prescription~\cite{Scimemi:2018xaf, Vladimirov:2019bfa}, the selection of the scale for TMDs is replaced by the selection of the equipotential line, for which we choose a line passing through the saddle point of the evolution field.
A TMD at a given set of scales ($\mu,\zeta$) is then given by
\begin{equation}
\tilde{f}_{q/p}(x,b_T;\mu,\zeta) = \tilde{f}_{q/p}(x,b_T) \left( \frac{\zeta}{\zeta_{\mu}(b_T)}\right)^{-\mathcal{D}(b_T;\mu)}\,,
    \label{e.final_TMD}
\end{equation}
where $\tilde{f}_{q/p}(x,b_T)$ is the so-called optimal TMD.  
This function is obtained by utilizing the OPE relation of TMDs and PDFs at small~$b_T$~\cite{Collins:2011zzd, Collins:2014jpa, Echevarria:2016scs, Luo:2020epw, Ebert:2020yqt, Ebert:2020qef, Moos:2020wvd} and modeling its nonperturbative large-$b_T$ behavior,
\begin{align}
\label{e.TMD.OPE}
  \tilde{f}_{q/p}(x,b_T) = & \!\! \sum_{f^\prime=q,\bar q,g} \left[C_{q/f^\prime}(x,b_T;\mu_{\scriptscriptstyle\rm OPE}) \otimes f_{f^\prime/p}(x;\mu_{\scriptscriptstyle\rm OPE}) \right]
  \nonumber \\
  &\hspace{1.2cm} \times
  f_{\rm NP}^q(x,b_T),
\end{align}
where $f_{f^\prime/p}$ is the collinear PDF for flavor $f^\prime$, and $f_{\rm NP}^q$ is a nonperturbative function that encodes large-$b_T$ behavior. 
The $\otimes$ symbol represents the convolution integral, $C \otimes f \equiv \int_x^1 (\df \xi/\xi)\, C(x/\xi)\, f(\xi)$.
Setting the scale
$\mu_{\scriptscriptstyle\rm OPE} = 2 e^{-\gamma_E}/b_T + 2$~GeV \cite{Moos:2023yfa} minimizes logarithmic corrections and avoids the Landau pole at large $b_T$.

To align with the NLO
treatment of the collinear observables in this analysis, the Wilson coefficients $C_{q/f^\prime}$ entering \eref{TMD.OPE} are taken at $\mathcal{O}(\alpha_s)$~\cite{Collins:2011zzd, Collins:2017oxh}.
Following the standard logarithmic counting~\cite{Bacchetta:2019sam, Boussarie:2023izj}, we implement NNLL accuracy for TMDs.
Specifically, we use $\Gamma$ at $\mathcal{O}(\alpha_s^3)$, and $\gamma_V$ and $\mathcal{D}_{\rm pert}$ at $\mathcal{O}(\alpha_s^2)$ precision, while the hard factor $\mathcal{H}^{{\mbox{\tiny \rm DY}}}_{q\bar q}$ in \eref{DYpTxsec} is at $\mathcal{O}(\alpha_s^2)$. 
We plan to extend the perturbative accuracy to higher orders in the future, as inclusion of NNLO order corrections can further improve the description of the collinear-sensitive data~\cite{NNPDF:2024nan}.

{\it Phenomenological framework --- }
The 
data used in this analysis include DIS from BCDMS~\cite{BCDMS:1989ggw}, NMC~\cite{NewMuon:1996fwh, NewMuon:1996uwk}, SLAC~\cite{Whitlow:1991uw}, Jefferson Lab~\cite{JeffersonLabE00-115:2009jll, Seely:2009gt, JeffersonLabHallATritium:2021usd, CLAS:2014jvt} and HERA~\cite{H1:2015ubc};
DY lepton-pair production from the Fermilab NuSea~\cite{NuSea:2001idv} and SeaQuest~\cite{SeaQuest:2021zxb} experiments;
$W^{\pm}$-lepton asymmetries from CMS~\cite{CMS:2011bet, CMS:2012ivw, CMS:2013pzl, CMS:2016qqr} and LHCb~\cite{LHCb:2014liz, LHCb:2015mad} at the LHC and STAR~\cite{STAR:2020vuq} at RHIC;
reconstructed $Z/\gamma^*$ and $W^{\pm}$ asymmetries from the Tevatron~\cite{D0:2013lql, CDF:2009cjw,  D0:2007djv, CDF:2010vek};
$W$+\,charm production~\cite{ATLAS:2014jkm, CMS:2013wql, CMS:2018dxg} from the LHC;
and jet production data from the Tevatron~\cite{PhysRevLett.101.062001, PhysRevD.75.092006} and RHIC~\cite{PhysRevLett.97.252001}.
We note that while state-of-the-art purely collinear analyses include $q_T$-integrated LHC $Z$-boson production data~\cite{Hou:2019efy, Bailey:2020ooq, NNPDF:2021njg}, our emphasis is on the inclusion of their $q_T$-dependent counterparts, which we describe using the TMD formalism. 
Our baseline PDF set does not include these high-precision $q_T$-integrated measurements and is therefore comparatively less constrained.
A relatively low cut on DIS data of $W^2 > 3.5$~GeV$^2$ allows sensitivity to quark distributions in the large-$x$ region.

For the processes described above we utilize the $\overline{\text{MS}}$ renormalization scheme, with the strong coupling $\alpha_s$ evolved numerically using the QCD $\beta$-functions with boundary condition $\alpha_s(M_Z) = 0.118$ at the $Z$-boson mass. 
The PDFs are evolved to NLL accuracy solving the DGLAP evolution equations~\cite{Dokshitzer:1977sg, Gribov:1972ri, Altarelli:1977zs,Sato:2016wqj} using Mellin-space techniques in the zero-mass variable flavor scheme.
The input scale $\mu_0$ is set to the charm quark mass, $m_c = 1.28$~GeV \cite{ParticleDataGroup:2024cfk}.
The collinear PDFs are parametrized at $\mu_0$ using the shape function
\begin{align}
    f(x;\mu_0) = N x^\alpha (1-x)^\beta (1 + \gamma \sqrt{x} + \delta x)\;,
    \label{eq.PDFparametrization}
\end{align}
which historically was motivated by Regge theory~\cite{Regge1959, Donnachie1992} for $\alpha$ and quark counting rules~\cite{BrodskyFarrar1973, Matveev1973} for $\beta$, although in modern QCD analyses these are treated as completely free parameters.
The shape-modifying factor with the parameters $\gamma$ and $\delta$ adds flexibility in the intermediate-$x$ region~\cite{supplemental}.
A total of 35 parameters are used for the $u_v$, $d_v$, $\bar{u}$, $\bar{d}$, $s$, $\bar{s}$, and $g$ flavors, while heavy quark distributions are generated through evolution.
As usual, the number and momentum sum rules are imposed to constrain the normalization parameters for the $u_v$, $d_v$, $s$ and $g$ PDFs.
We additionally include contributions from higher twist effects, target mass corrections, and off-shell effects that are important in describing the high-$x$ DIS data~\cite{Cocuzza:2021rfn, Cocuzza:2026zoy}.

In this analysis, 
we use $q_T$-dependent DY data from both fixed target and collider experiments.
The fixed target inputs are from the E288 experiment~\cite{Ito:1980ev} taken with 200, 300, and 400~GeV proton beams on a copper (Cu) target, the E605 experiment~\cite{Moreno:1990sf} taken on a Cu target, and the E772 experiment~\cite{E772:1994cpf} using a deuterium target. 
For the high-energy collider datasets, we analyze absolute cross sections, including $p \bar{p}$ data from the Tevatron integrated in rapidity, comprising two datasets from CDF~\cite{CDF:1999bpw, CDF:2012brb} at 1.8 and 1.96~TeV and D0~\cite{D0:1999jba} at 1.8~TeV.
We also use LHC $pp$ data taken around the $Z$ boson peak: ATLAS~\cite{ATLAS:2015iiu} at 8~TeV and CMS~\cite{CMS:2011wyd} at 13~TeV binned in rapidity, and LHCb data~\cite{LHCb:2015okr, LHCb:2015mad, LHCb:2021huf} at 7, 8, and 13~TeV integrated over large rapidity, $2 \leq y \leq 4.5$.
Finally, $pp$ data from the STAR 510~GeV~\cite{STAR:2023jwh} and PHENIX 200~GeV runs~\cite{PHENIX:2018dwt} at RHIC are also included.

For consistency between our theoretical formalism and the experimental data, we impose kinematic cuts for the TMD-sensitive data such that $q_T/Q < 0.2$, and at least one of $q_T^2 / Q^2 < 2 \sigma$ or $q_T < 10$~GeV is satisfied, where $\sigma$ is the experimental relative uncorrelated uncertainty~\cite{Moos:2023yfa, Moos:2025sal}.
This yields 436 points for the $q_T$-dependent DY data, which together with the 4279 points for collinear observables, amounts to a total dataset of 4715 points.

For the parametrization of the nonperturbative ingredients of the TMDs, we follow Ref.~\cite{Moos:2023yfa} and take
\begin{align}
    f_{\rm NP}^q(x,b_T) ={\rm cosh}\left[\left( \lambda_1^q (1-x) + \lambda_2^q x \right)b_T\right]^{-1} \; ,
\end{align}
where $\lambda_{1,2}^q$ are free parameters for $u$, $d$, $\bar u$, $\bar d$ and $q_{\rm sea} = s = \bar{s} = c = \bar{c} = b = \bar{b}$.
The nonperturbative part of the CS kernel is parametrized as 
\begin{equation}
\mathcal{D}_{\rm NP}(b_T) = b_T b^* \left[ c_0 + c_1 \ln \left({{b^*}}/{B_{\rm NP}}\right)\right]\,,
\label{eq.DNP}
\end{equation}
where $c_0$ and $c_1$ are free parameters.
This parametrization is motivated by the expected exponential decay of TMDs and a linear dependence of the Collins-Soper kernel at large $b_T$~\cite{Collins:2013zsa, Collins:2015dfa, supplemental}.

Since the fixed-target data involve nuclei, we relate the TMDs for the nucleus to the bound proton and neutron TMDs by
    $\tilde f_{q/{A}} \equiv (Z/A)\, \tilde f_{q/{p}} + (1-Z/A)\, \tilde f_{q/{n}}$ 
for a nucleus with mass number $A$ and atomic number $Z$.
We do not include additional nuclear effects in this analysis; nuclear TMD effects have previously been studied~\cite{Alrashed:2022jlx, Barry:2023qqh} and will be incorporated in future work.
In all we have 12 parameters for the TMD sector, which together with the 35 collinear PDF shape parameters and the 12 higher-twist and off-shell parameters, gives a total of 59 parameters that need to be inferred from the data.
This constitutes a computational challenge that requires a high performance computing effort due to the lengthy calculation involved, particularly for the OPE.

\begin{table}[b] 
\caption{$\chi^2/N_{\rm pts}$ and $Z$-score values for the  PDF only (\texttt{baseline PDF}) and TMD only (\texttt{baseline TMD+LHC}) analyses compared with the combined TMD$+$PDF analysis ($\rm JAM25_{PDF+TMD}$).\\}
\centering
\begin{tabular}{ l r c | c }
\hline
Process & $N_{\rm pts}$~ & \multicolumn{2}{c}{$\chi^2/N_{\rm pts}~(Z$-score)} \\
\hline
&  &  &  \\
\multicolumn{2}{l}{\bf Collinear} & PDF only & TMD+PDF \\ \hline
DIS -- fixed target 
& 2501~ & 1.02~(+0.82) & 1.02~(+0.59)
\\
\phantom{DIS} -- HERA 
& 1185~ & 1.27~(+6.01) & 1.27~(+6.11) 
\\
DY 
&  205~ & 1.27~(+2.54) & 1.26~(+2.50) \\
$W$-$\ell$ asymmetry  
&   70~ & 0.80~($-$1.20) & 0.78~($-$1.35) 
\\
$W$ asymmetry 
&   27~ & 1.12~(+0.51) & 1.13~(+0.53) 
\\
$Z$\! rapidity 
&   56~ & 1.09~(+0.55) & 1.11~(+0.60) 
\\
jets 
&  198~ & 1.00~(+0.00) & 0.98~($-$0.13)
\\
$W\!+\!c$ 
&   37~ & 0.65~($-$1.66) & 0.62~($-$1.84) 
\\ \hline
Total collinear  
& 4279~ & 1.10~(+4.31) & 1.09~(+4.14) \\
\hline 
&  &  &  \\
\multicolumn{2}{l}{\bf TMD} & TMD only & TMD+PDF \\ \hline
DY -- E288,\! E605,\! E772 
& 224~ & 1.36 (+3.44) & 1.31~(+3.02) 
\\
\phantom{DY} -- CDF, D0 
&  80~ & 1.06~(+0.45) & 1.11~(+0.71) 
\\
\phantom{DY} -- STAR,\! PHENIX 
&  12~ & 1.15~(+0.47) & 1.20~(+0.60) 
\\
\phantom{DY} -- ATLAS 
&  30~ & 2.07~(+3.29) & 1.78~(+2.55) 
\\
\phantom{DY} -- CMS 
&  64~ & 1.18~(+1.03) & 0.92~($-$0.39) 
\\
\phantom{DY} -- LHCb 
&  26~ & 0.53~($-$1.97) & 0.50~($-$2.12) 
\\ \hline
Total TMD  
& 436~ & 1.27~(+3.72) & 1.20~(+2.76) 
\\
\hline 
&  &  &  \\
\bf{Total}  
& \bf{4715}\, &  & ~\bf{1.10}~(+4.79) 
\\
\hline
\end{tabular}
\label{t.chi2comparison}
\end{table}

For our global analysis we use the Bayesian Monte Carlo methodology developed by the JAM Collaboration~\cite{Jimenez-Delgado:2013sma, Jimenez-Delgado:2013boa, Jimenez-Delgado:2014xza, Sato:2016tuz, Sato:2016wqj, Ethier:2017zbq, Lin:2017stx, Barry:2018ort, Sato:2019yez, Cammarota:2020qcw, Bringewatt:2020ixn, Moffat:2021dji, Adamiak:2021ppq, Cao:2021aci, Cocuzza:2021rfn, Zhou:2021llj, Barry:2021osv, Cocuzza:2021cbi, Zhou:2022wzm, Boglione:2022gpv, Cocuzza:2022jye, Barry:2022aix, Gamberg:2022kdb, Cocuzza:2026zoy}, generating $\mathcal{O}(1\,000)$ replicas to accumulate sufficient statistics to compute approximate credible intervals (CIs) for the collinear and TMD PDFs.
We also filter replicas with large posterior $Z$-scores, where $Z = \Phi^{-1}(p) \equiv \sqrt{2} {\rm erf}^{-1}(2p-1)$, with the $p$ value of the resulting $\chi^2$ computed relative to the expected $\chi^2$ distribution.

{\it Results --- }%
We perform the Bayesian Monte Carlo analysis for four different scenarios:
\begin{enumerate}
    \item 
    \underline{\texttt{baseline PDF}}:~collinear PDFs inferred from collinear observables only;
    \item 
    \underline{\texttt{baseline TMD}}:~TMDs inferred from the $q_T$-dependent fixed target DY, Tevatron, and RHIC data, with PDF replicas fixed to the baseline PDF set.  This allows PDF uncertainties to be propagated into the extracted TMDs;
    \item 
    \underline{\texttt{baseline TMD\,+\,LHC}}:~TMDs inferred using in addition LHC TMD-sensitive data, with TMD priors taken from baseline TMD and PDFs fixed from baseline PDF.  This allows us to assess the impact of LHC data;
    \item 
    \underline{$\bf JAM25_{PDF+TMD}$}:~full simultaneous PDF and TMD inference set, combining all collinear-sensitive and TMD-sensitive data considered in this work. 
\end{enumerate}

The resulting agreement with the data is shown in Table~\ref{t.chi2comparison}, where the quoted values of $\chi^2$ and $Z$-scores are computed using the average theory over the ensemble of replicas.
The combined analysis leads to an improvement in the description of the experimental measurements, especially for the $q_T$-dependent DY data where the $Z$-score reduces by $\approx$ 1 unit from 3.72 to 2.76, while the collinear-sensitive datasets remain stable with a $Z$-score reduction of only 0.17.
The improvement in $\chi^2/N_{\rm pts}$ is greatest for ATLAS data (from 2.07 to 1.78) and CMS (from 1.18 to 0.92).
We stress that the ATLAS data are extremely precise, and all modern extractions fitting only DY data have relatively large $\chi^2$ values~\cite{Bacchetta:2022awv, Moos:2023yfa, Bacchetta:2025ara}.

\begin{figure}[t] 
    \centering
    \includegraphics[width=0.9\columnwidth]{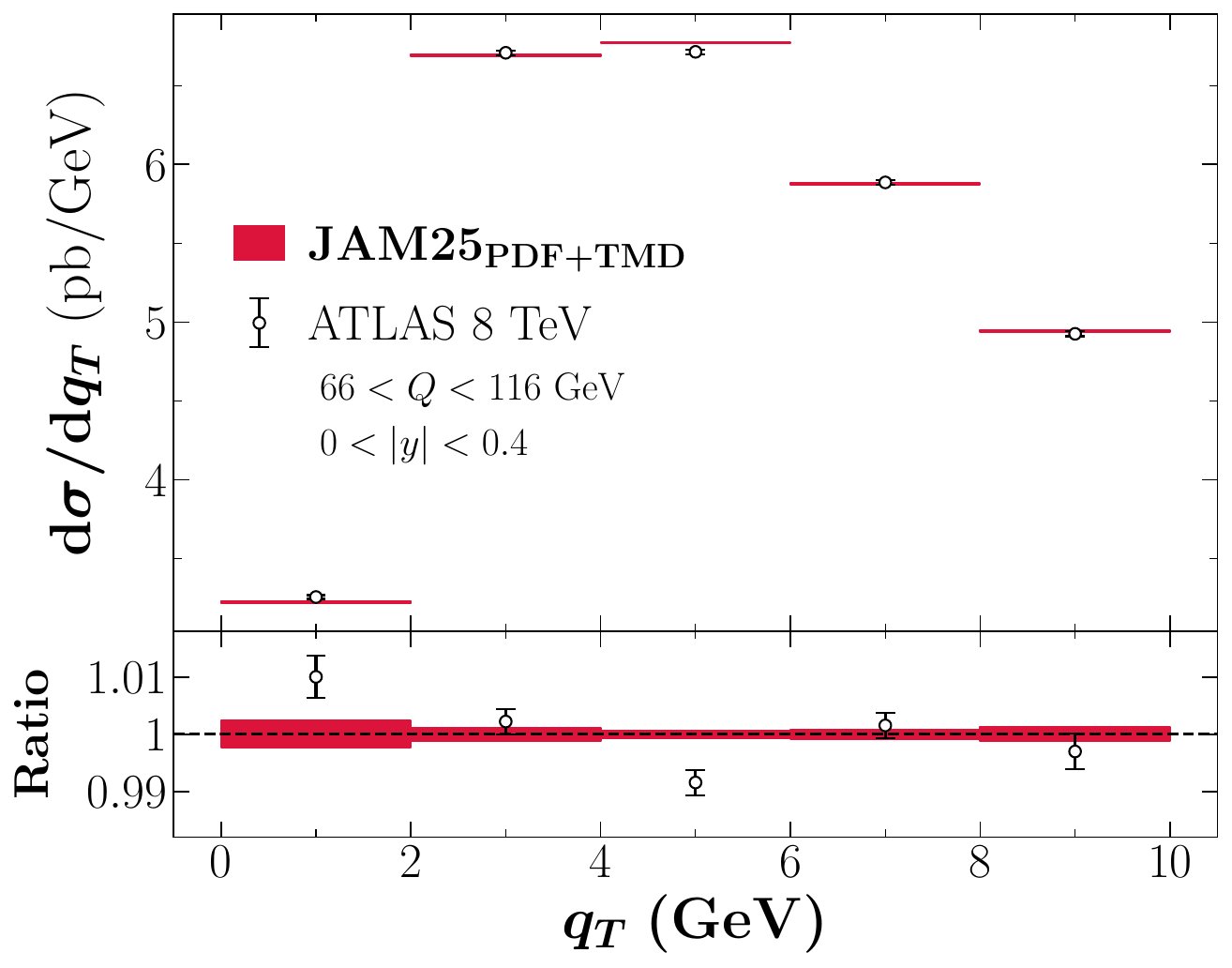}
    \vspace{-0.25cm}
    \caption{
    Comparison of a representative bin of the ATLAS 8~TeV data with the $\rm JAM25_{PDF+TMD}$ analysis at the 68\%~CI.}
    \label{f.data_theory}
\end{figure}

In Fig.~\ref{f.data_theory}, we show the data and theory comparison for the $0 \leq |y| \leq 0.4$ bin of the ATLAS data, as well as the data to theory ratio.
We note the sub-percent precision of the data and our ability to describe the data to within 1\%.
We additionally achieve a high degree of consistency of normalization coefficients for all $q_T$-dependent DY datasets, with all fitted normalization factors being within $1\sigma$ of the reported normalization uncertainty, except in the case of ATLAS and CMS where the agreement is within $3\sigma$.

\begin{figure}[t] 
    \centering
    \includegraphics[width=1\columnwidth]{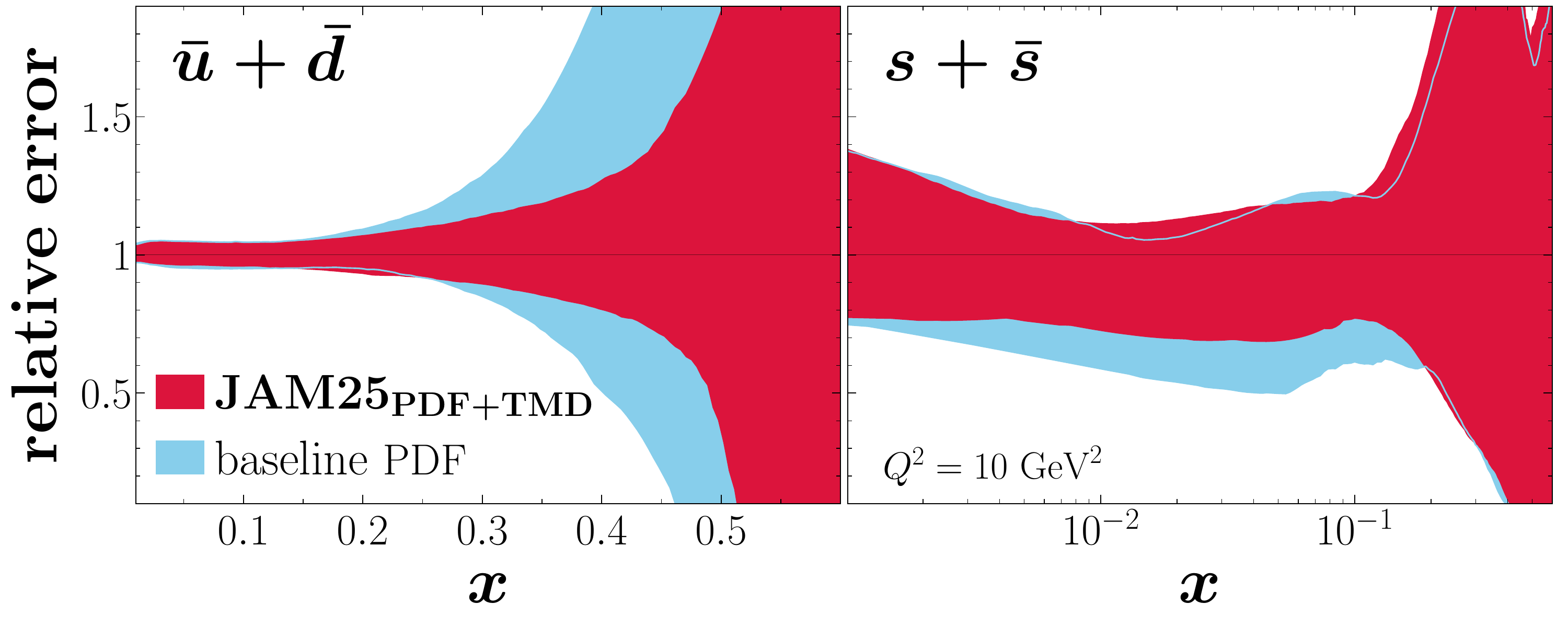}
    \vspace*{-0.6cm}
    \caption{Comparison of relative errors for collinear $\bar{u} + \bar{d}$ (left panel) and $s + \bar s$ (right panel) PDFs between the baseline PDF (blue) and $\rm JAM25_{PDF+TMD}$ (red) analyses  at $Q^2=10$~GeV$^2$ for 95\%~CIs. 
    }
    \label{f.PDF}
\end{figure}

In Fig.~\ref{f.PDF}, we present the $\bar{u}+\bar{d}$ and $s+\bar{s}$ PDFs from the \texttt{baseline PDF} and $\rm JAM25_{PDF+TMD}$ analyses as errors relative to the mean value of the combined PDF+TMD analysis.
We observe a significant reduction in the uncertainties for the $\bar{u}+\bar{d}$ PDFs for $x \gtrsim 0.3$, largely influenced by the fixed target $q_T$-dependent DY data, and a modest reduction in the strange quark sector for $10^{-3}\leq x\leq 0.2$ due to the LHC data.
While the magnitude of the light sea decreases at large $x$, since the relative uncertainty increases, we have verified that the reduction persists for the absolute uncertainty of $x(\bar{u}+\bar{d})$.
The reduction in uncertainty of the $s+\bar{s}$ PDFs is consistent with the inclusion of the collinear-sensitive LHC data as observed in Ref.~\cite{NNPDF:2017mvq}.
The comparable reduction obtained from the $q_T$-integrated LHC data indicates that this constraint on the strange PDF is not unique to the TMD-sensitive data. 
Rather, the similar effect observed in our PDF+TMD analysis as in the purely collinear analysis \cite{NNPDF:2017mvq} supports the consistency of these two theoretical frameworks.

The improvement of the uncertainties on the extracted TMDs is demonstrated in~\fref{TMD}, where we show $\tilde f_{q/p}(x,b_T)$ for $q=u$, $\bar{u}$, and $s$  at $Q=10$~GeV and $x=0.2$, 0.4, and 0.1, respectively, where the impact of the PDF+TMD combined analysis is most prominent.
The constraints on the $u$ and $s$ quark TMDs come largely from the LHC data, as shown by the uncertainty reduction from the yellow to green bands.
However, these bands for the $\bar{u}$ TMD are rather similar, indicating  that it is largely constrained by the fixed target $q_T$-dependent DY data, with the LHC data playing a minor role.
Our simultaneous analysis of TMDs and PDFs has the strongest impact on the $\bar{u}$ and $s$ quark TMDs, particularly at $b_T \lesssim 1$~GeV$^{-1}$.
Interestingly, although the combined analysis significantly impacts both sectors, we find only weak statistical correlation between the PDF and TMD parameters (see Supplemental Material for details~\cite{supplemental}).

In \fref{CS}, we show the extracted CS kernel from Eq.~\eqref{e.cs} at $Q=10$~GeV.
One can see that the inclusion of the LHC data significantly constrains its values below $b_T=1.5$~GeV$^{-1}$. 
In addition, the extractions of the CS kernel from the $\rm JAM25_{PDF+TMD}$ and baseline TMD+LHC analyses are largely compatible.
This shows that there is a significant decorrelation between the CS kernel and PDFs.
The inset of \fref{CS} compares the CS kernel at $Q=2$~GeV to the lattice QCD results from Ref.~\cite{Avkhadiev:2024mgd}.
Our results are in excellent agreement with the lattice QCD data, as well as other phenomenological extractions~\cite{Bacchetta:2025ara, Moos:2025sal, Moos:2023yfa, Bacchetta:2022awv, Bacchetta:2019sam} (not shown in \fref{CS}).

\begin{figure*}[t] 
   \centering
    \includegraphics[width=2.09\columnwidth]{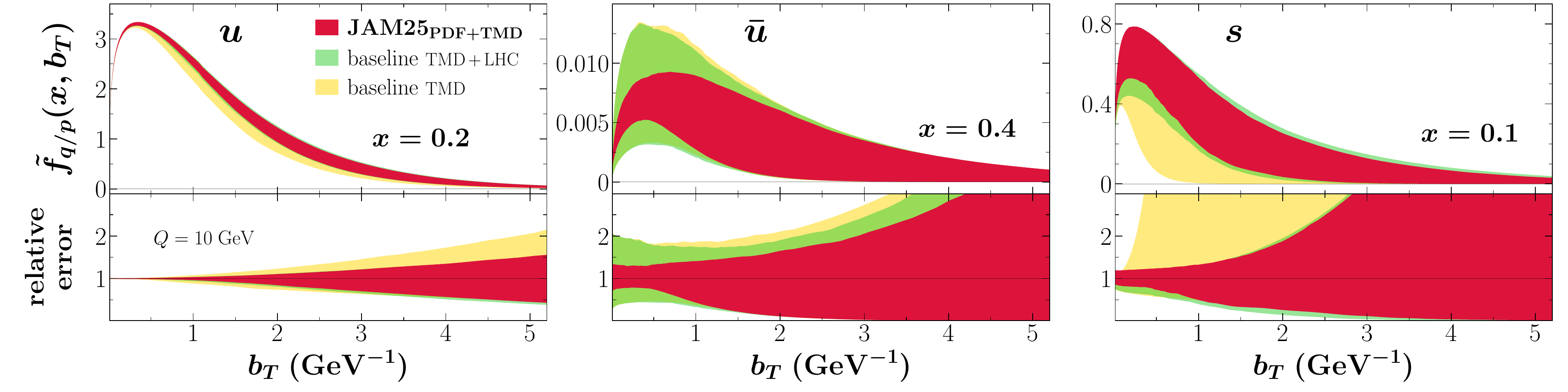}
    \vspace*{-0.3cm}
    \caption{TMDs $\tilde f_{q/p}(x,b_T)$ (upper panels) and ratios to the mean values of $\tilde f_{q/p}$ (bottom panels) as functions of $b_T$ for $u$, $\bar u$, and $s$ quarks at fixed $x=0.2$, 0.4, and 0.1, respectively, and $Q=10$~GeV. 
    The $\rm JAM25_{PDF+TMD}$ results (red) are compared with the \texttt{baseline TMD} (yellow) and \texttt{baseline TMD+LHC} (green) analyses for 95\%~CIs.
    }
    \label{f.TMD}
\end{figure*}

\begin{figure}[h] 
   \centering
    \includegraphics[width=0.93\columnwidth]{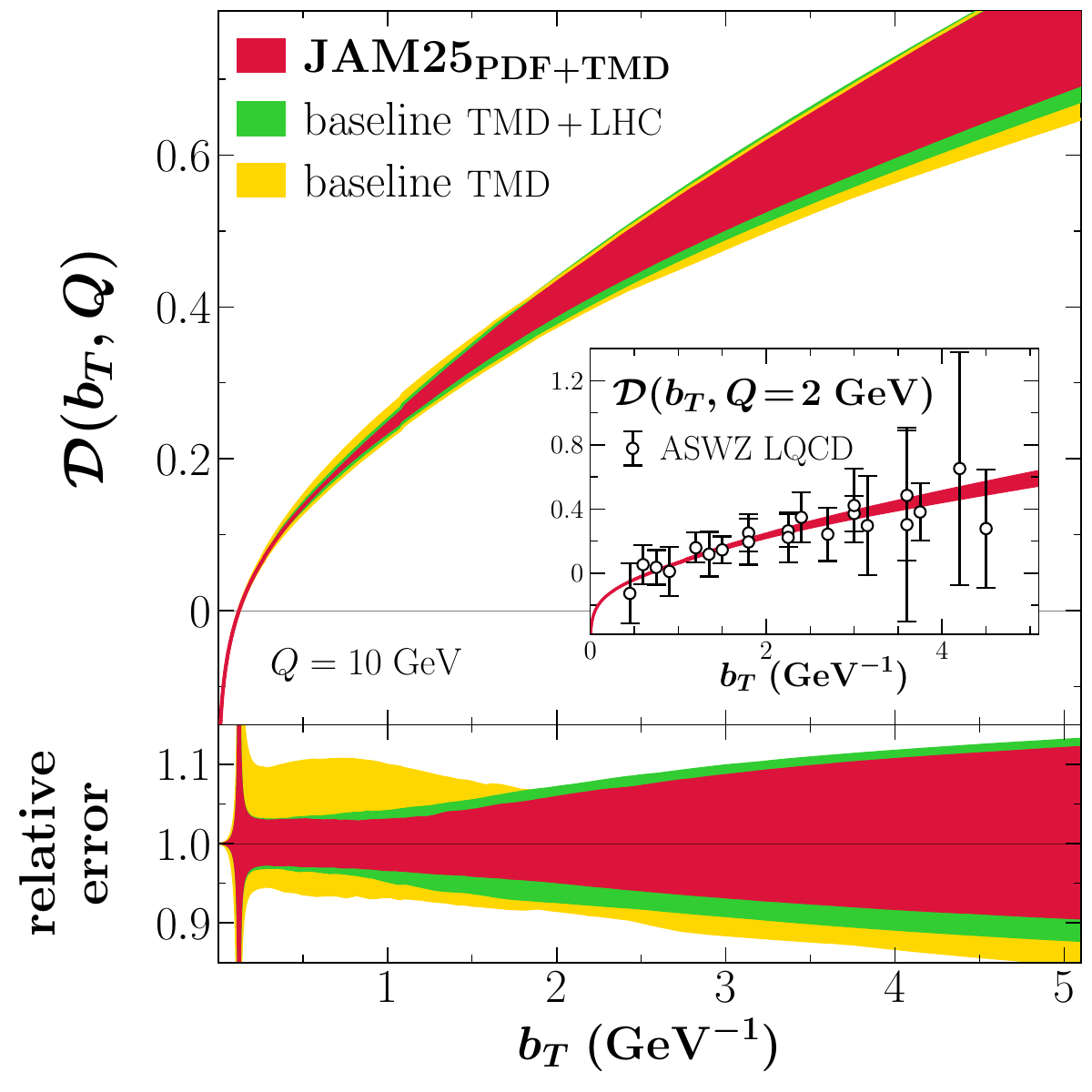}
    \vspace*{-0.3cm}
    \caption{Collins-Soper kernels ${\mathcal D}(b_T,Q)$ at $Q=10$~GeV at 95\%~CIs, with notations as in \fref{TMD}.
    The inset compares the 68\%~CI $\rm JAM25_{PDF+TMD}$ results with lattice QCD (ASWZ LQCD) data at $Q=2$~GeV~\cite{Avkhadiev:2024mgd}.
    (The lattice data were originally presented in Ref.~\cite{Avkhadiev:2024mgd} as $\gamma_q \equiv -2 \mathcal{D}$.)
    }
    \label{f.CS}
\end{figure}

{\it Outlook ---} 
Our simultaneous extraction of proton TMDs and PDFs, obtained from a large collection of observables ranging from low-energy fixed target experiments to high-energy LHC data, opens a new frontier for QCD studies of nucleon structure.
The observed reductions in PDF and TMD uncertainties, especially in the sea-quark sector, together with improvements in the Collins–Soper kernel, highlight the potential for a unified phenomenological treatment of TMD and collinear physics in future simultaneous global QCD analyses.

In the future, we will extend our phenomenological framework to include large transverse momentum data that are also sensitive to collinear PDFs~\cite{Cao:2021aci, Camarda:2025lbt}, and perform precision studies of the QCD strong coupling~\cite{Cerci:2023uhu}.
Our ultimate goal is to carry out a comprehensive phenomenological study that additionally includes observables sensitive to hadronization effects via semi-inclusive annihilation in $e^+e^-$ and semi-inclusive DIS reactions from existing Belle, Jefferson Lab, HERMES, and COMPASS experiments, preparing for the future Electron-Ion Collider~\cite{AbdulKhalek:2021gbh}.
This work paves the way to test and validate the universality of reconstructed nonperturbative hadron structures that emerge in QCD from GeV to TeV scales.\\

{\it Acknowledgments} ---  
We thank Yong Zhao for providing the lattice QCD data for the CS kernel.
This work was supported by the U.S.~Department of Energy, Office of Science, Office of Nuclear Physics, contract No.~DE-AC02-06CH11357 (P.B. and E.M.), the Scientific Discovery through Advanced Computing (SciDAC) award {\it Femtoscale Imaging of Nuclei using Exascale Platforms} (P.B.), by the U.S. Department of Energy contract No.~DE-AC05-06OR23177, under which Jefferson Science Associates, LLC operates Jefferson Lab (T.A., W.M., A.P., J.Q., N.S.), contract No.~DE-SC0026320 (L.G.), the National Science Foundation under Grants No.~PHY-2308567 (D.P.), and No.~PHY-2310031, No.~PHY-2335114 (A.P.).
The work of N.S. and R.W. was supported by the DOE, Office of Science, Office of Nuclear Physics in the Early Career Program.
This project is supported by the European Union Horizon research Marie Skłodowska-Curie Actions – Staff Exchanges, HORIZON-MSCA-2023-SE-01-101182937-HeI, DOI: 10.3030/101182937 (A.V., A.P., J.Q., N.S.).
The work of A.V. is funded by the \textit{Atracci\'on de Talento Investigador} program of the Comunidad de Madrid (Spain) No. 2020-T1/TIC-20204, and the grant ``Europa Excelencia'' No. EUR2023-143460 funded by MCIN/AEI/10.13039/501100011033/ by the Spanish Ministerio de Ciencias y Innovaci\'on. 
This work has benefited by interactions within the Quark-Gluon Tomography (QGT) Topical Collaboration funded by the U.S. Department of Energy, Office of Science, Office of Nuclear Physics with Award DE-SC0023646. \\ \\

\bibliography{references_no_arxiv_for_published}

\clearpage
\onecolumngrid

\setcounter{equation}{0}
\setcounter{figure}{0}
\setcounter{table}{0}
\setcounter{section}{0}
\setcounter{subsection}{0}

\renewcommand{\theequation}{S\arabic{equation}}
\renewcommand{\thefigure}{S\arabic{figure}}
\renewcommand{\thetable}{S\arabic{table}}
\renewcommand{\thesection}{S\arabic{section}}
\renewcommand{\thesubsection}{S\arabic{section}.\arabic{subsection}}

\setcounter{secnumdepth}{2}

\begin{center}
{\bfseries \large Supplemental Material}\\[0.75em]

\end{center}

\vspace{0.5em}

This following provides technical details of the fitting methodology and analysis presented in the Letter.

\section{Parametrizations of collinear and TMD distributions}
\label{s.parametrization}

The extraction of parton distributions in global QCD analysis constitutes an inverse problem and is typically made tractable by adopting parametric functional forms that facilitate their determination. 
While in principle any sufficiently flexible parametrization could be employed, in this work we adopt traditional forms of the collinear PDF parametrization, historically motivated by Regge theory and spectator counting rules, according to Eq.~(6) in the main text.
In particular, Regge theory suggests a small-$x$ behavior proportional to $x^\alpha$~\cite{Regge1959, Donnachie1992}, while spectator counting rules have been used to motivate a large-$x$ dependence proportional to $(1-x)^\beta$~\cite{BrodskyFarrar1973, Matveev1973}.
To provide additional flexibility in the intermediate-$x$ region, we include a shape-modifying factor $(1+\gamma\sqrt{x}+\delta x)$, leading to the form in Eq.~(6).
We emphasize that this parametrization only specifies the shape at the input scale $\mu_0$; the scale dependence of the distributions is determined entirely by the DGLAP evolution equations.

Our analysis employs a total of 35 PDF shape parameters. 
After imposing the relevant sum rules, the valence $u$- and $d$-quark and gluon distributions are each described by four free parameters. 
The sea-quark sector is parametrized using separate light-sea (``ls'') and strange-sea (``ss'') contributions. 
For the light-sea quarks,
\begin{subequations}
\begin{align}
\bar{u} &= \bar{u}_0+q_{\rm ls}\, , \\
\bar{d} &=\bar{d}_0+q_{\rm ls}\, ,
\end{align}
\end{subequations}
where the common component $q_{\rm ls}$ is shared between the $\bar u$ and $\bar d$ distributions. 
The light-quark sea thus contains a total of 15 parameters, with five parameters assigned to each of $\bar u_0$, $\bar d_0$, and $q_{\rm ls}$.
For the strange-quark sector, we analogously have,
\begin{subequations}
\begin{align}
s&=s_0+q_{\rm ss}\, , \\
\bar{s}&=\bar{s}_0+q_{\rm ss}\, .
\end{align}
\end{subequations}
Because of the limited experimental constraints available for strange quarks, we set $\gamma=\delta=0$ for the $s$ and $\bar s$ PDFs.
Consequently, the functions $s_0$, $\bar s_0$, and $q_{\rm ss}$ contain two (because of the strange number rule), three, and three free parameters, respectively, yielding a total of eight strange-sea parameters.

Using the same functional form, six additional parameters are used to describe higher-twist contributions to the proton and neutron $F_2$ structure functions by setting $\beta=\gamma=0$. 
A further six parameters account for off-shell corrections in nuclear DIS, corresponding to two distributions, each parametrized using the same functional form as Eq.~(6) in the main text with $\gamma=\delta=0$. 
Additional details about the higher-twist and off-shell parametrizations can be found in Refs.~\cite{Cocuzza:2021rfn,Cocuzza:2026zoy}.

The TMD parametrization in Eqs.~(7) and (8) of the main text is motivated by the analysis of Ref.~\cite{Moos:2023yfa}, which presented a successful description of DY data. 
Additional guidance comes from the expected exponential suppression of TMDs and the approximately linear behavior of the Collins-Soper kernel at large $b_T$ values~\cite{Collins:2013zsa,Collins:2015dfa}.

\subsection{Correlations}

\begin{figure}[t]
    \centering
    \includegraphics[width=0.95\linewidth]{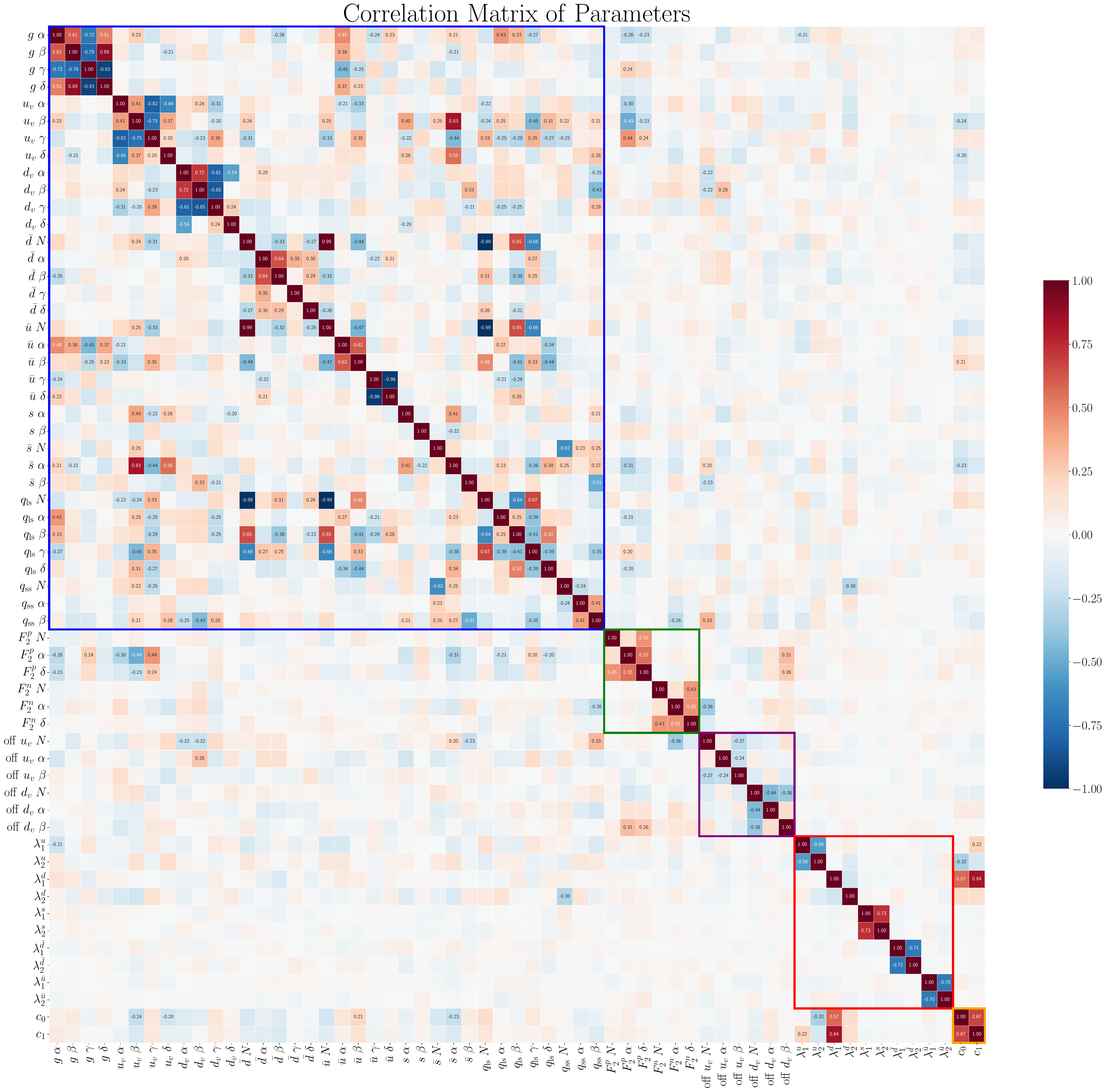}
    \caption{Correlation matrix for all fitted parameters used in this analysis, including those for collinear PDFs (blue box), higher-twist contributions (green), nuclear off-shell corrections for DIS observables (purple), nonperturbative intrinsic TMDs (red), and nonperturbative components of the Collins-Soper kernel (orange).}
    \label{f.corrmat}
\end{figure}

In Fig.~\ref{f.corrmat}, we present the correlation matrix for all fitted parameters obtained from the full ensemble of Monte Carlo replicas. 
The correlation coefficient between parameters $a_i$ and $a_j$ is defined as
\begin{equation}
\rho_{ij}
=
\frac{\langle a_i a_j\rangle-\langle a_i\rangle\langle a_j\rangle}
{\sigma_i\,\sigma_j},
\end{equation}
where $\sigma_i^2=\langle a_i^2\rangle-\langle a_i\rangle^2$, and the averages are taken over the replica ensemble.
No strong correlations are observed between the TMD and collinear PDF parameters, with the corresponding correlation coefficients $|\rho_{ij}| < 0.2$ for most cross sector parameter pairs.

In the figure we denote the level of correlation for those entries having $|\rho_{ij}| > 0.2$. 
Note that even small modifications in the small-$b_T$ region can propagate to large $b_T$ through the OPE, which fixes the normalization of the TMD. 
Consequently, even relatively weak correlations between the two sectors can have a non-negligible phenomenological impact. 
The relationship between the collinear PDF parameters and the nonperturbative TMD parameters is highly nonlinear.
There are nontrivial components, such as DGLAP evolution and convolution with the Wilson coefficients for the OPE, that are multiplicatively combined with the nonperturbative TMD parameters which then undergo a Fourier transform.
As such, the correlations among parameters measure the linearity between two random variables, such that the relative impacts between the two sectors are not evident from the correlations alone.
Indeed, we find that the simultaneous analysis improves the determination of both the collinear PDFs and the TMD distributions.

\subsection{Additional parametric form for TMDs}

\begin{figure}[ht]
    \centering
    \includegraphics[width=0.8\linewidth]{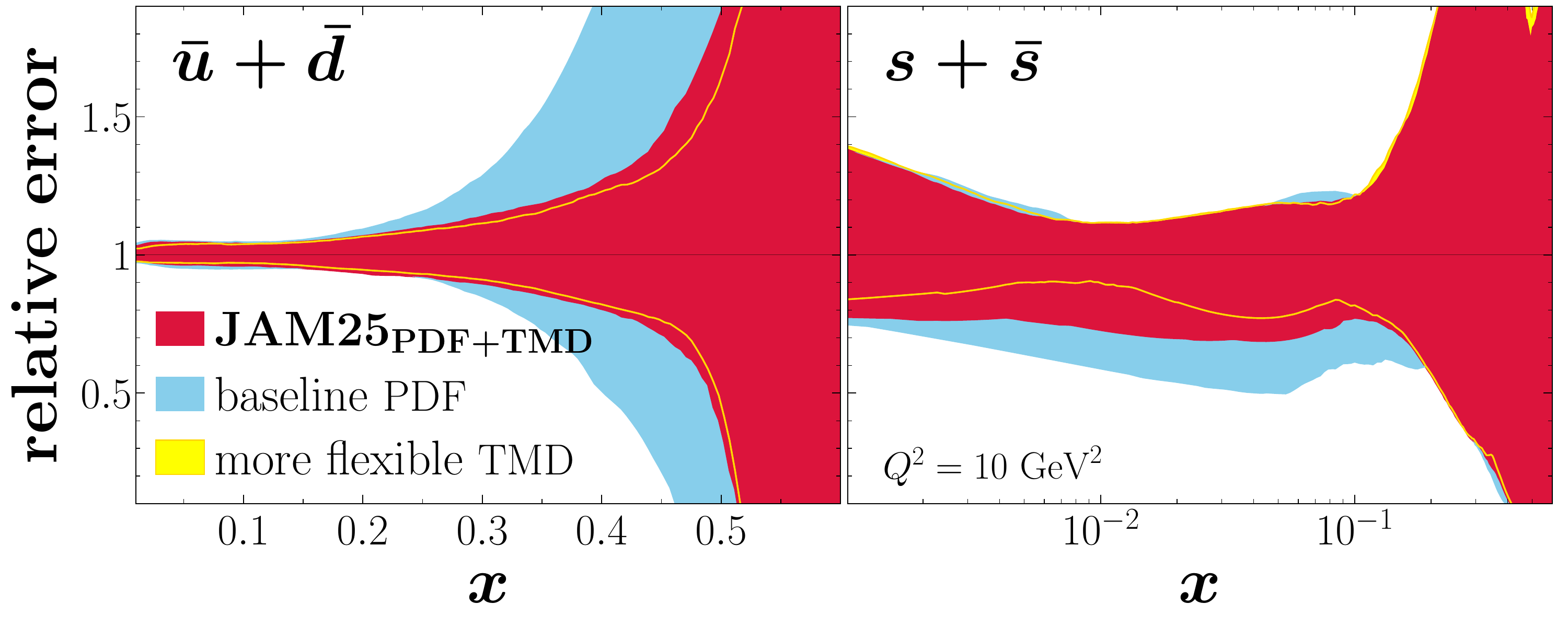}
    \caption{As in Fig.~2 in the main text, but now including a fit with a more flexible parametrization of the nonperturbative TMD (yellow bands).}
    \label{f.more_flexible_TMD}
\end{figure}

In addition to the nonperturbative TMD parametrization adopted in this work, we performed an alternative fit using a more flexible parametrization of the form,
\begin{align}
    f_{\rm NP}^q(x,b_T) = {\rm cosh}\left[\left( \lambda_1^q (1-x)^{\lambda_3^q} + \lambda_2^q x \right)b_T\right]^{-1} \; ,
    \label{eq.more_flexible_TMD}
\end{align}
which was previously employed in Ref.~\cite{Moos:2025sal}.
In the analysis presented in this Letter, the parameter $\lambda_3^q$ was fixed to unity for all quark flavors. 
Here, we relax this constraint and treat $\lambda_3^q$ as a free parameter for each parametrized flavor. 
This provides a useful test of the sensitivity of the extracted collinear PDFs to the assumed functional form of the nonperturbative TMD.
The resulting PDFs are compared in Fig.~\ref{f.more_flexible_TMD}.

We find that the inclusion of TMD-sensitive data continues to produce visible modifications of the collinear PDFs even when a more flexible nonperturbative TMD parametrization is employed. 
This indicates that the impact of the TMD measurements is not an artifact of the specific functional form adopted in the Letter, but rather reflects genuine constraints provided by the data.
The final results reported in the Letter, however, were obtained using the parametrization of Eq.~(7).

\section{Methodology Choices}

\subsection{Closure Tests}

To further validate the robustness of the fitting methodology introduced in this Letter, we performed a series of closure tests. 
A single Monte Carlo replica was selected as the ground truth, defining the PDFs, TMDs, and the higher-twist and off-shell parameters described in Sec.~\ref{s.parametrization}.
Using this parameter set, we generated pseudodata for all observables included in the $\mathrm{JAM25}_{\mathrm{PDF+TMD}}$ analysis.

Three complementary closure tests were performed. 
In the first test, the collinear sector parameters were fixed to their true values, while only the TMD parameters were fitted to the TMD observables. 
In the second test, the TMD parameters were fixed and the collinear sector parameters were fitted using the complete pseudodata set, including both collinear and TMD observables. 
Finally, in the third test, all parameters were released and simultaneously fitted to the full pseudodata set.

\begin{figure*}[th]
    \centering
    \includegraphics[width=1.\linewidth]{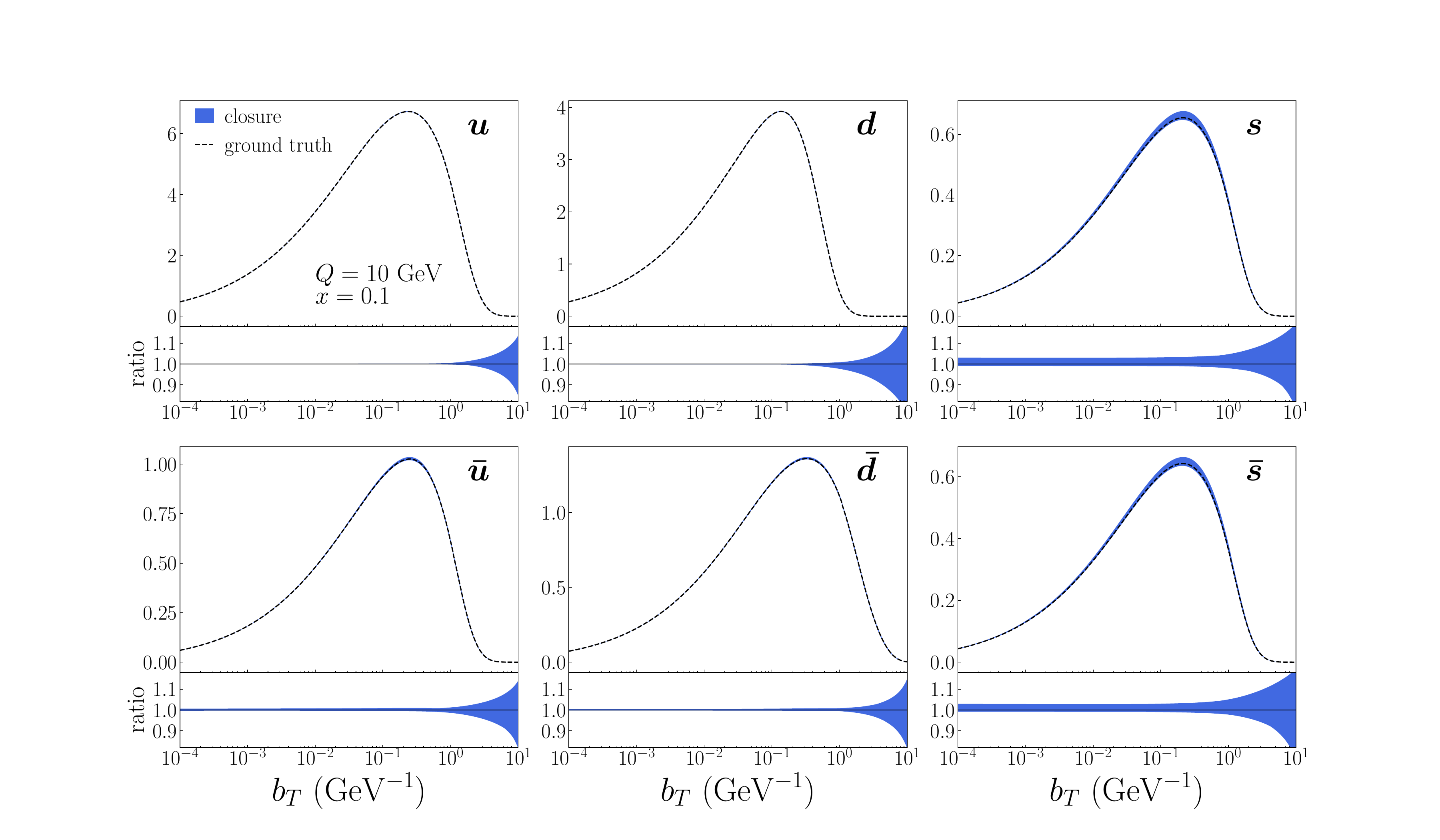}\\ (a)
    \includegraphics[width=1.\linewidth]{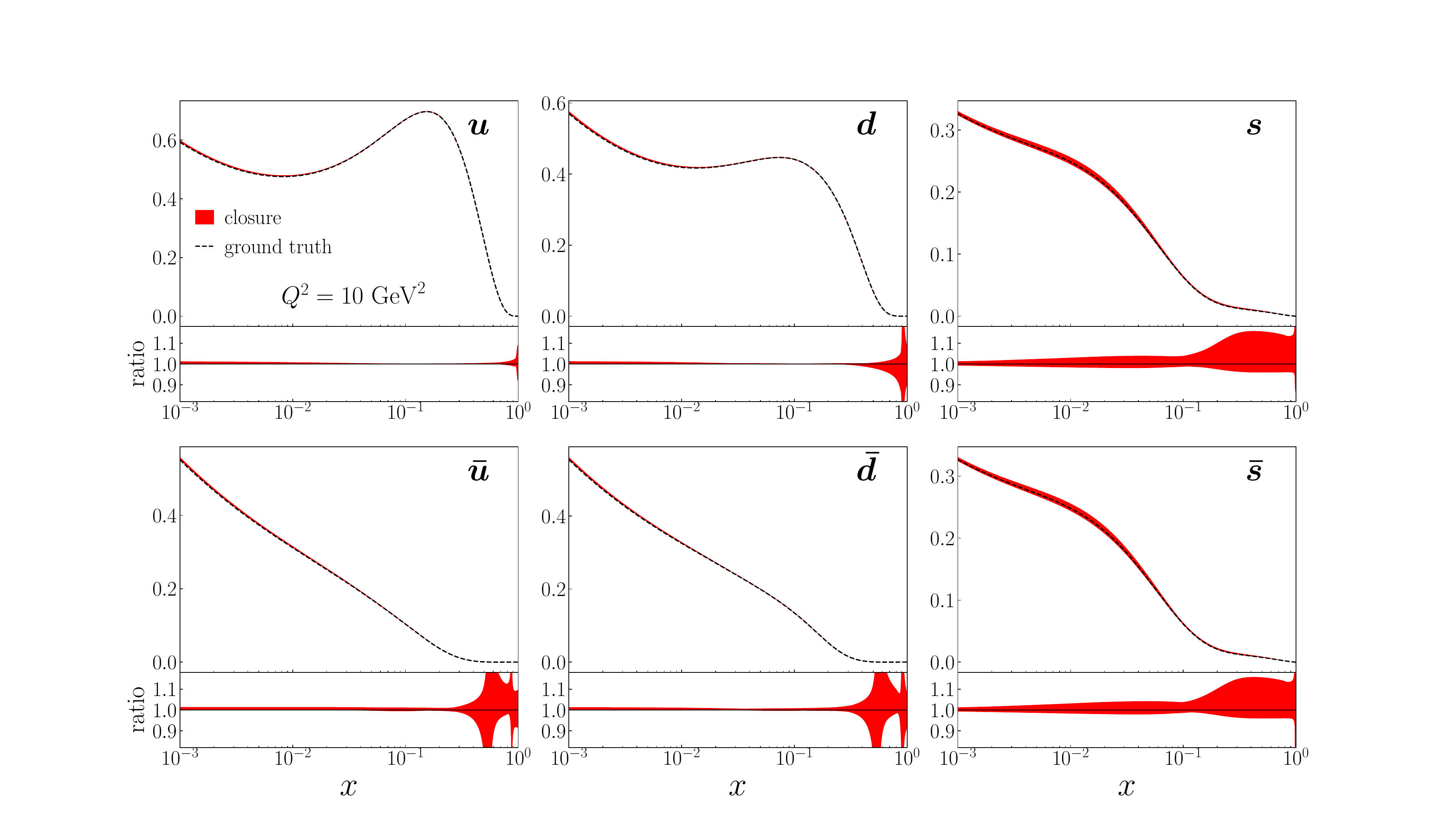}\\ (b)
    \caption{Results for the $u$, $d$, $s$, $\bar u$, $\bar d$ and $\bar s$ flavors of the closure test 3, obtained by creating pseudodata for all datasets and fitting both TMDs and PDFs:
    {\bf Top two panels (a):} TMD PDFs versus $b_T$ at $x=0.1$ and $Q^2=10$~GeV$^2$;
    {\bf Bottom two panels (b):} Collinear PDFs versus $x$ at $Q^2=10~{\rm GeV}^2$.
    For each flavor the subpanels show the ratio of the fitted result to the generating replica.}
    \label{f.closuretest3_tmd}
\end{figure*}

These tests provide a direct assessment of the ability of the fitting framework to recover the underlying parameter values and correctly propagate information between the collinear and TMD sectors.
When fitted to pseudodata generated without bootstrap fluctuations, all three closure tests exactly recovered the generating replica, yielding $\chi^2=0$ and reproducing the corresponding PDFs and TMDs. 
This confirms the internal consistency of the fitting framework and minimization procedure.

To assess the impact of statistical fluctuations, we also performed closure tests using 10 bootstrapped pseudodata replicas. 
The results of the third closure test, in which all PDF and TMD parameters are simultaneously fitted to the full pseudodata set, are shown in Fig.~\ref{f.closuretest3_tmd}(a) for the TMD PDFs versus $b_T$ at $x=0.1$ and $Q=10$~GeV, and in Fig.~\ref{f.closuretest3_tmd}(b) for the PDFs versus $x$ at $Q^2=10~{\rm GeV}^2$, for each of the $u$, $d$, $s$, $\bar u$, $\bar d$ and $\bar s$ flavors.
For each flavor, the panels display the extracted distributions, while the subpanels show the ratio of the fitted result to the generating replica.

As can be seen from Fig.~\ref{f.closuretest3_tmd}, the underlying replica is reproduced within the uncertainties obtained from the ensemble of 10 pseudodata replicas for all fitted distributions. 
Similar agreement is observed for the first and second closure tests.
This demonstrates that the fitting methodology reliably recovers the input PDFs and TMDs, and correctly propagates uncertainties in both the individual and simultaneous fits.

\subsection{Scanning $q_T/Q$}

\begin{figure}[t]
    \centering
    \includegraphics[width=0.6\linewidth]{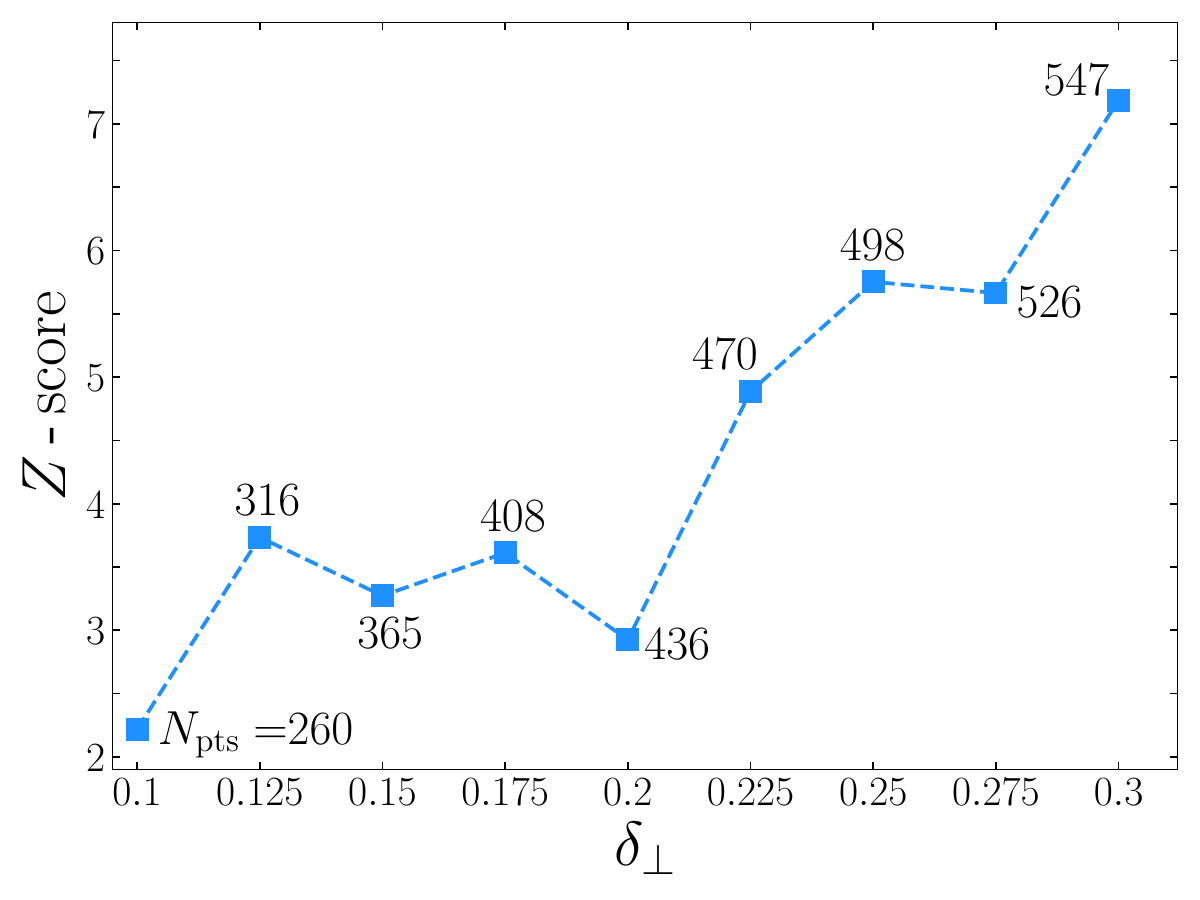}
    \caption{$Z$-score values as a function of the upper cut $\delta_\perp$ on $q_T/Q$. The label on each point in $\delta_\perp$ indicates the number of data points, $N_{\rm pts}$, for the TMD-sensitive datasets. The dashed lines between the points are included to guide the eye.}
    \label{f.Z_delta_perp}
\end{figure}

To assess the choice of the kinematic cut defining the region of validity of TMD factorization, $q_T/Q < q_T^{ \rm max}/Q \equiv \delta_\perp$, we performed a series of single fits without bootstrap resampling, scanning values of $\delta_\perp$ between 0.1 and 0.3. 
For each value of $\delta_\perp$, a full simultaneous PDF+TMD fit was carried out.
In this study, we additionally imposed the cut $q_T^2/Q^2 < 2\sigma$ or $q_T < 10$~GeV.
These restrictions are necessary to suppress contributions from $\mathcal{O}(q_T/Q)$ power corrections, which can become significant at the level of precision reached by modern LHC measurements and are not included in our leading-twist description.

The results of the $\delta_\perp$ scan are shown in Fig.~\ref{f.Z_delta_perp}, where we display the corresponding $Z$-scores for the TMD-sensitive datasets. 
As expected, the $Z$-score generally increases with increasing $\delta_\perp$, reflecting the gradual deterioration of the fit quality as data with larger transverse momenta are included. 
The choice $\delta_\perp=0.2$ provides a favorable balance between fit quality and statistical coverage, yielding a $Z$-score of approximately 3 while retaining 436 TMD-sensitive data points in the analysis.

Increasing $\delta_\perp$ further leads to a noticeable degradation in the description of the TMD-sensitive data. 
This may indicate the onset of effects beyond the range of applicability of the present TMD-only framework, including both power corrections and the increasing relevance of the fixed-order collinear contribution. 
A complete description of the broader $q_T$ spectrum would require a matched $W+Y$ treatment~\cite{Collins:1984kg}, where the $W$ term corresponds to the TMD factorization employed here and the $Y$ term accounts for the large-$q_T$ collinear contribution, and will be interesting to consider in a future analysis.

\section{Impact of fixed target and LHC datasets}

\begin{figure}[t]
    \centering
    \includegraphics[width=0.7\linewidth]{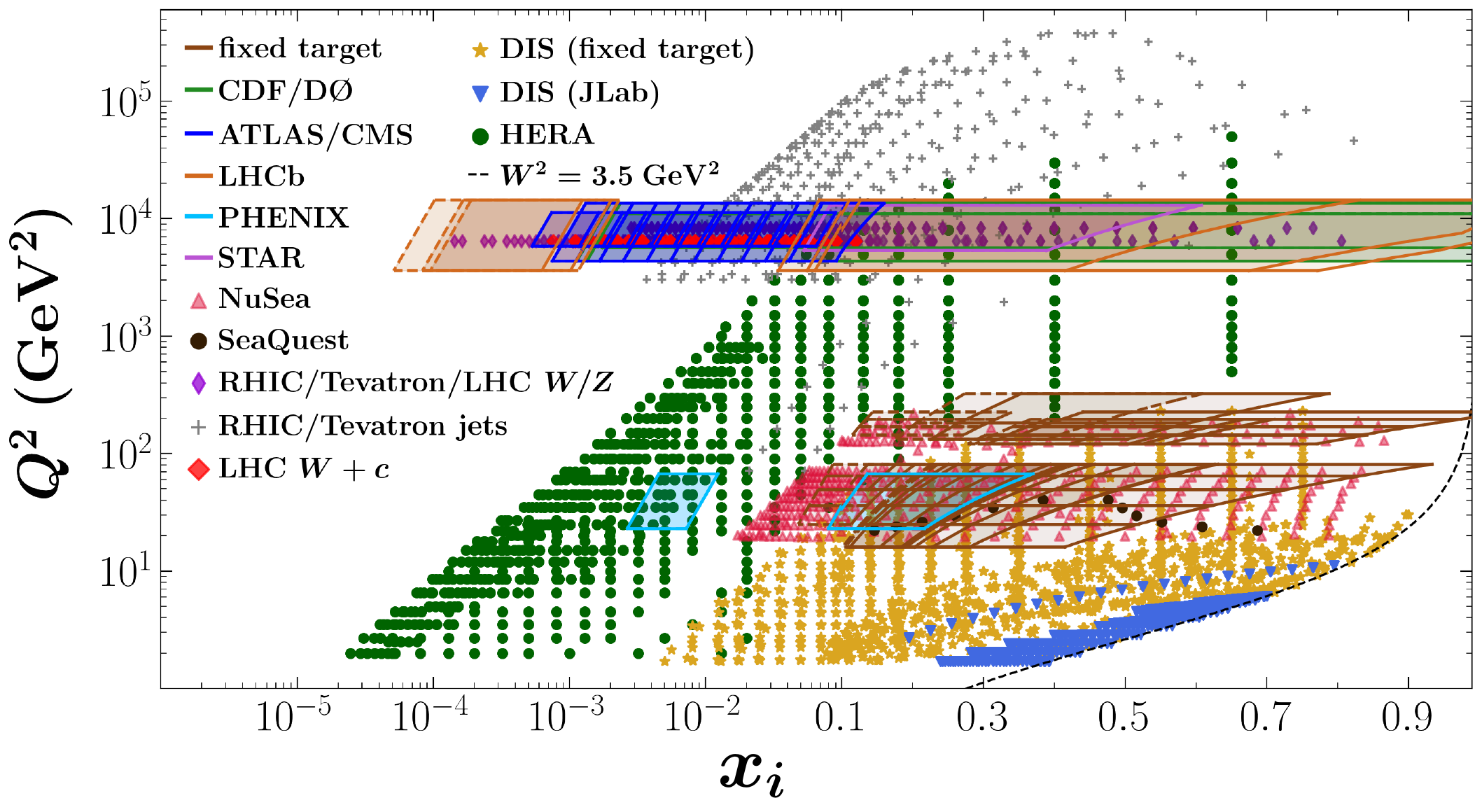}
    \caption{Kinematics of ($x,Q^2$) for the data used in this analysis. The $q_T$-dependent data are represented by shaded regions, as they are presented in ranges of $x$ and $Q^2$.}
    \label{f.kinematics}
\end{figure}

Our analysis indicates that in the TMD sector the fixed-target measurements have the largest impact on the isoscalar sea-quark combination $\bar{u}+\bar{d}$, whereas the LHC data provide the strongest constraints on the strange-quark distributions. 
To understand the origin of this behavior, we first examine the kinematic coverage in the $(x,Q^2)$ plane shown in Fig.~\ref{f.kinematics}.

The fixed-target $q_T$-dependent DY data primarily probe the region $0.1 \lesssim x \lesssim 1$ (indicated by the darker brown shaded areas in Fig.~\ref{f.kinematics}). 
By contrast, most of the $q_T$-dependent LHC measurements are sensitive to the region $10^{-3} \lesssim x \lesssim 0.1$. 
The main exception is the LHCb data, which probe larger values of $x$ due to their coverage of the forward rapidity region, $2 \le y \le 4.5$.
These kinematic considerations suggest that the fixed-target DY measurements primarily constrain the moderate- and large-$x$ quark and antiquark distributions, while the LHC data are predominantly sensitive to the small-$x$ sea-quark sector. 
Consequently, one expects the fixed-target data to have the greatest impact on the light-sea distributions, whereas the LHC measurements provide enhanced sensitivity to the strange-quark distributions compared to the fixed-target data alone.

\begin{figure}[b]
    \centering
    \includegraphics[width=0.8\linewidth]{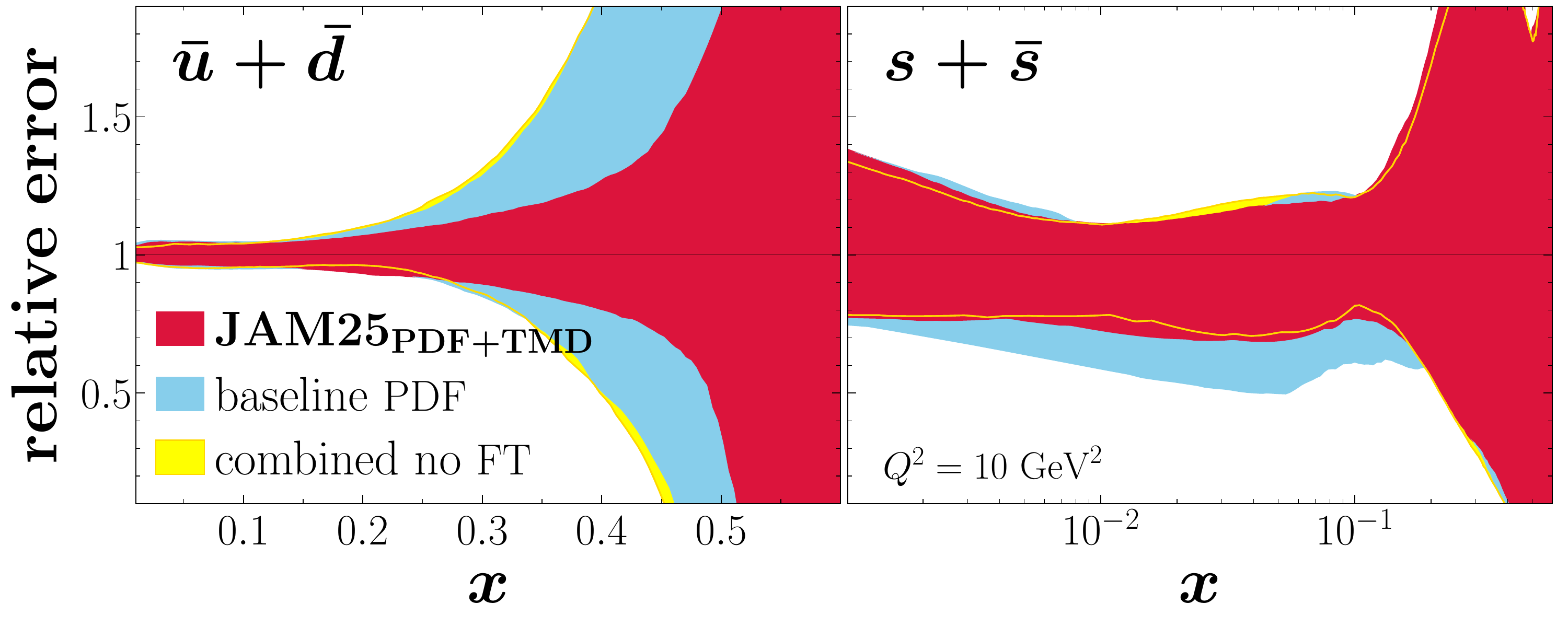}
    \caption{As in Fig.~2 of the main text, but removing the fixed-target (FT) data (yellow bands).}
    \label{f.noFT}
\end{figure}

To more directly assess the impact of the fixed-target $q_T$-dependent DY measurements, we performed an additional simultaneous PDF+TMD fit (using the full set of collinear observables together with the Tevatron, RHIC, and LHC TMD-sensitive data) in which these data were excluded.
The resulting PDFs are shown in Fig.~\ref{f.noFT}. 
The removal of the fixed-target DY data leads to a substantial reduction in the impact of the TMD sector on the light-sea combination $\bar{u}+\bar{d}$. 
Indeed, the resulting distributions are nearly indistinguishable from those obtained in the baseline collinear-only fit, indicating that the sensitivity to $\bar{u}+\bar{d}$ observed in the full analysis originates primarily from the fixed-target measurements.
In contrast, the extracted $s+\bar{s}$ distributions remain in close agreement with the final $\mathrm{JAM25}_{\mathrm{PDF+TMD}}$ results. 
This demonstrates that the constraints on the strange-quark sector arise predominantly from the collider TMD-sensitive measurements, which are included in the fit.

\begin{figure}[t]
    \centering
    \vspace*{-0.3cm}
    \includegraphics[width=0.54\linewidth]{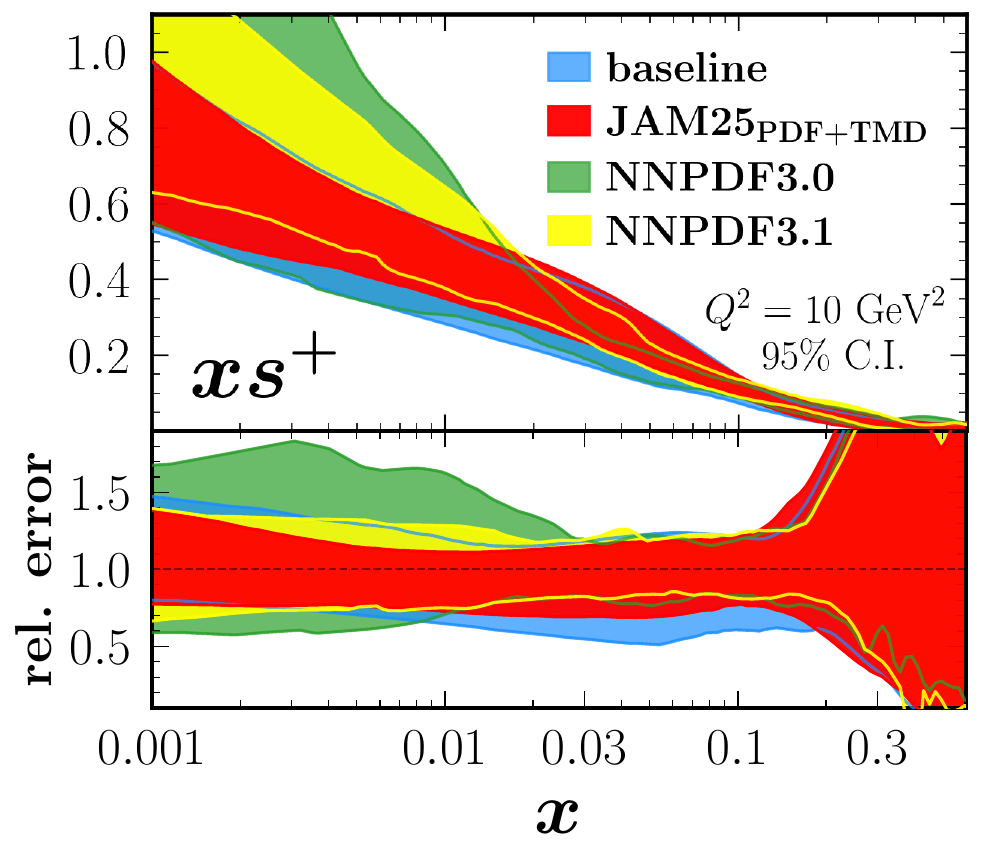}
    \caption{Comparison of the $s^+ \equiv s + \bar s$ collinear PDFs for our baseline and $\mathrm{JAM25}_{\mathrm{PDF+TMD}}$ fits with the NNPDF3.0~\cite{NNPDF:2014otw} and NNPDF3.1~\cite{NNPDF:2017mvq} sets.}
    \label{f.spluscomparison}
\end{figure}

To assess the impact of including $q_T$-integrated LHC data on the strange-quark distributions, we compare the NNPDF3.0~\cite{NNPDF:2014otw} and NNPDF3.1~\cite{NNPDF:2017mvq} determinations, where the latter incorporates the newest $q_T$-integrated DY LHC measurements that are absent in the former.
As shown in Fig.~\ref{f.spluscomparison}, the inclusion of these data leads to a reduction in the uncertainty of the strange-quark combination $s^+ \equiv s + \bar{s}$ that is comparable to the reduction observed in our simultaneous PDF+TMD analysis.

This comparison suggests that the constraints provided by the $q_T$-differential LHC measurements included in our fit are broadly consistent with those obtained from the corresponding $q_T$-integrated LHC data. 
The observed reduction of the strange-quark uncertainty therefore supports the interpretation that both analyses are sensitive to similar features of the proton's strange-quark content. 
Consequently, as Fig.~\ref{f.spluscomparison} indicates, the strange-quark distributions obtained in our simultaneous PDF+TMD fit remain highly compatible with those extracted in state-of-the-art collinear analyses such as NNPDF3.1~\cite{NNPDF:2017mvq}. 

\vspace*{-0.5cm}

\end{document}